\documentclass[11pt]{article}
\usepackage{amsmath,amssymb}

\makeatletter
\@addtoreset{equation}{section}
\makeatother

\def\crampest{\medmuskip = 1mu plus 1mu minus 1mu}
\def\uncramp{\medmuskip = 4mu plus 2mu minus 4mu}
\def\ben{\begin{equation}}
\def\een{\end{equation}}
\def\half{{\textstyle{1\over2}}}

 \let\m=\mu \let\n=\nu  \let\p=\pi

\let\C=\Chi

\def\nn{\nonumber} \def\bd{\begin{document}} \def\ed{\end{document}}
\def\ds{\documentstyle} \let\fr=\frac \let\bl=\bigl \let\br=\bigr
\let\Br=\Bigr \let\Bl=\Bigl
\let\bm=\bibitem
\let\na=\nabla
\let\pa=\partial \let\ov=\overline
\newcommand{\be}{\begin{equation}}
\newcommand{\ee}{\end{equation}}
\def\ba{\begin{array}}
\def\ea{\end{array}}
\def\ft#1#2{{\textstyle{\frac{\scriptstyle #1}{\scriptstyle #2} } }}
\def\fft#1#2{{\frac{#1}{#2}}}
\def\del{\partial}
\def\vp{\varphi}
\def\sst#1{{\scriptscriptstyle #1}}
\def\oneone{\rlap 1\mkern4mu{\rm l}}
\def\td{\tilde}
\def\wtd{\widetilde}
\def\ie{{\it i.e.\ }}
\def\dalemb#1#2{{\vbox{\hrule height .#2pt
        \hbox{\vrule width.#2pt height#1pt \kern#1pt
                \vrule width.#2pt}
        \hrule height.#2pt}}}
\def\square{\mathord{\dalemb{6.8}{7}\hbox{\hskip1pt}}}
\newcommand{\ho}[1]{$\, ^{#1}$}
\newcommand{\hoch}[1]{$\, ^{#1}$}
\newcommand{\bea}{\setlength\arraycolsep{2pt} \begin{eqnarray}}
\newcommand{\eea}{\end{eqnarray}}
\newcommand{\ra}{\rightarrow}
\newcommand{\lra}{\longrightarrow}
\newcommand{\Lra}{\Leftrightarrow}
\newcommand{\bp}{\tilde \beta^\prime}
\newcommand{\tr}{{\rm tr} }
\newcommand{\Tr}{{\rm Tr} }
\def\0{{\sst{(0)}}}
\def\1{{\sst{(1)}}}
\def\2{{\sst{(2)}}}
\def\3{{\sst{(3)}}}
\def\4{{\sst{(4)}}}
\def\5{{\sst{(5)}}}
\def\6{{\sst{(6)}}}
\def\7{{\sst{(7)}}}
\def\8{{\sst{(8)}}}
\def\m{{\sst{(m)}}}
\def\n{{\sst{(n)}}}
\def\cA{{{\cal A}}}
\def\cB{{{\cal B}}}
\def\cF{{{\cal F}}}
\def\cG{{{\cal G}}}
\def\cH{{{\cal H}}}
\def\tV{\widetilde V}
\def\tW{\widetilde W}
\def\tH{\widetilde H}
\def\tE{\widetilde E}
\def\tF{\widetilde F}
\def\tA{\widetilde A}
\def\im{{{\rm i}}}
\def\tY{{{\wtd Y}}}
\def\ep{{\epsilon}}
\def\vep{{\varepsilon}}
\def\bD{{{\bar D}}}
\def\R{{{\mathbb R}}}
\def\C{{{\mathbb C}}}
\def\H{{{\mathbb H}}}
\def\CP{{{\mathbb C}{\mathbb P}}}
\def\RP{{{\mathbb R}{\mathbb P}}}
\def\Z{{{\mathbb Z}}}
\def\bA{{{\mathbb A}}}
\def\bB{{{\mathbb B}}}
\def\bC{{{\mathbb C}}}
\def\bD{{{\mathbb D}}}
\def\bE{{{\mathbb E}}}

\def\bZ{{{\bf Z}}}
\def\bP{{{\bf P}}}
\def\bM{{{\bf M}}}
\def\bG{{{\bf G}}}
\def\bstarG{{\bf {G^\star}}}
\def\starF{{F^\star}}
\def\starG{{G^\star}}
\def\starZ{{Z^\star}}
\def\bstarZ{{\bf {Z^\star}}}
\def\startheta{{\theta^\star}}
\def\turk{{\.{I}n\"on\"u\ }}

\def\Re{{{\frak{Re}}}}
\def\Im{{{\frak{Im}}}}
\def\cosec{{\,\hbox{cosec}\,}}
\def\Gm{{\Gamma_{\!\! -}}}
\def\Gp{{\Gamma_{\!\! +}}}
\def\stan{{standard }}
\def\nonstan{{supernumerary }}
\def\p{{\partial}}
\def\kdel#1{{\fft{\del}{\del#1}}}

\def\bog{{Bogomolny }}
\def\om{{\omega}}

\newcommand{\nnr}{\nonumber \\}
\newcommand{\pd}{\partial}
\newcommand{\ud}{\textrm{d}}
\newcommand{\dTH}{T^{\prime \, 0}_\textrm{H}}
\newcommand{\dOi}{\Omega^{\prime \, 0}_i}

\newcommand{\tamphys}{\it George and Cynthia Woods Mitchell  Institute
for Fundamental Physics and Astronomy,\\
Texas A\&M University, College Station, TX 77843-4242, USA}

\newcommand{\auth}{ G.W. Gibbons\hoch{\ddagger},
Joaquim Gomis\hoch{\diamondsuit} and
C.N. Pope\hoch{\dagger,\ddagger} }

\begin{document}

\begin{flushright}
\hfill{ICCUB-09-227,
MIFP-09-42, UB-ECM-PF-09/23, DAMTP-2009-56\ \ \ \ \ \ }\\
\end{flushright}

\begin{center}
{\large {\bf Deforming the Maxwell-Sim Algebra}}

\vspace{15pt}
\auth

\vspace{10pt}
\hoch{\dagger}{\tamphys}

\vspace{10pt}


\vspace{10pt}

\hoch{\ddagger}{\it  DAMTP, Centre for Mathematical Sciences,
 Cambridge University,\\  Wilberforce Road, Cambridge CB3 OWA, UK}

\vspace{10pt}

\hoch{\diamondsuit}{\it Departament d'Estructura i Constituents de
la Mat\`eria and ICCUB,\\ Universitat de Barcelona, Diagonal 647,
08028 Barcelona}


\vspace{30pt}

\underline{ABSTRACT}

\end{center}

   The Maxwell algebra is a non-central extension of the Poincar\'e
algebra, in which the momentum generators no longer commute, but satisfy
$[P_\mu,P_\nu]=Z_{\mu\nu}$. The charges $Z_{\mu\nu}$ commute with the
momenta, and transform tensorially under the action of the angular momentum
generators.  If one constructs an action for a massive particle, invariant
under these symmetries, one finds that it satisfies the equations of
motion of a charged particle interacting with a constant electromagnetic
field via the Lorentz force.  In this paper, we explore the analogous
constructions where one starts instead with the ISim subalgebra of
Poincar\'e, this
being the symmetry algebra of Very Special Relativity. It admits an analogous
non-central extension, and we find that a particle action invariant under
this  Maxwell-Sim algebra again describes a particle subject to
the ordinary Lorentz force.  One can also deform the ISim algebra to
DISim$_b$, where $b$ is a non-trivial dimensionless parameter.  We find that
the motion described by an action invariant under the corresponding
Maxwell-DISim algebra is that of a particle interacting via a Finslerian
modification of the Lorentz force. In an appendix is it shown that
the  DISim$_b$ algebra is isomorphic to the extended Schr\"odinger algebra
with $b=\frac{1}{1-z}$.

\vspace{15pt}

\thispagestyle{empty}

\pagebreak
\setcounter{page}{1}

\tableofcontents

\addtocontents{toc}{\protect\setcounter{tocdepth}{2}}


\newpage

\section{Introduction}

A popular line of thought in theoretical physics is to start with a
Lie algebra $\frak{g}$ or Lie group $G$, and then to construct from
it the space or spacetime  in which physical objects, for
examples $p$-branes, move. Typically  the spaces or spacetimes are
cosets $G/H$.
The dynamics of $p$-branes is then described as a map from the
$(p+1)$-dimensional world volume into $G/H$
\cite{lees,dirac,Nambu:1970b,Goto:1971ce,Hughes:1986dn}. In the case
of point particles, the dynamics is often thought of as geodesic
motion,  or some modification thereof by ``forces,'' such as the
Lorentz force on electrically charges particles in electromagnetism,
 with respect to a metric on $G/H$ that is invariant
under the left action of $G$ on $G/H$. More generally, one is
interested in invariant Lagrangians $L(x,v)$ on the tangent space
$T(G/H)$, or Hamiltonians $H(x,p)$ on the cotangent space
$T^\star(G/H)$.  For a recent statement of this viewpoint
in the context of quantum field theory see \cite{Schwarz}.

  An alternative construction of a $p$-brane action in the space $G/H$ is to
consider the quotient $(G/H)/K$,  where $K$ is the stabilizer of the $p$-brane.
The action of lowest order in derivatives is obtained by considering the
pull-back to the world-volume of a $(p+1)$-form invariant under $K$
\cite{Gauntlett:1989qe} (see also \cite{Gomis:2006xw}, where one can
find further references). The action contains extra Goldstone fields
associated with the broken ``rotations.'' In order to make
contact with the geometrical Lagrangian $L(x,v)$, we should
eliminate the extra fields by their non-dynamical equations of motion, or more
generally, by the inverse Higgs mechanism \cite{Ivanov:1975zq}.

    An early example of this programme followed
the discovery of the three congruence geometries;
hyperbolic or Lobachevsky space $H^3$, Euclidean space ${\Bbb E}^3$,
and spherical space $S^3$.  Helmholtz characterised
these three possibilities physically in terms of
axioms of the {\it free mobility of rigid bodies} \cite{Helmholtz}.
Such bodies permit
rotations about any point in space, and translations
to any point in space. Thus he demanded that $H=SO(3)$
and that $G$ act transitively on $G/H$.  He arrived, after some
additional arguments, at the three possibilities
\bea
G=SO(3,1): \qquad  G/H &=&H^3\,,\\
G=E(3): \qquad   G/H &=& {\Bbb E} ^3  \,,\\
G=SO(4): \qquad   G/H &=& S^3   \,.
\eea

   An equivalent way of looking at this
is to say that the configuration space $Q$
of a rigid body with one point fixed
admits a  simply-transitive left  action by  $SO(3)$, and may thus
be identified with $SO(3)$. Free motion of a rigid body
is given by geodesic motion on $SO(3)$ with respect
to a left-invariant metric. If the body moves in ordinary
Euclidean space, $Q$ is enlarged  to become
the Euclidean group $E(3)$. If the body moves in an inviscid fluid,
conservation of momentum  and angular momentum
will still  hold and the metric is then given by a general left-invariant
metric on  $E(3)$. Correspondingly, geodesic motion on $SO(3,1)$ or $SO(4)$
with respect to a left-invariant metric gives the motion
of a rigid body moving in a fluid in   $H^3$
or $S^3$ respectively.

   A slightly different strand of thought begins with the observation
(originally due to Lambert \cite{lam})  that passing to $S^3$ or $H^3$
introduces a new parameter into physics: the radius of curvature.
This new parameter is associated with the fact that the translations
in $SO3,1)$ or $SO(4)$  no longer commute. Expressed mathematically,
the Lie algebras $\frak{so}(3,1)$ and $\frak{so}(4)$ are {\it
continuous deformations} of $\frak{e}(3)$, and this suggests that in
seeking new physical laws, a fruitful procedure is to look for
continuous deformations of existing laws.  In algebraic terms, this
translates into looking for continuous deformations of the Lie
algebra $\frak{g}$ that one begins with. Given that one  has
introduced a new physical parameter whose magnitude is arbitrary, it
is natural to enquire whether it might be {\sl time-dependent}.
In the case of spatial curvature, just such a suggestion was made by
Calinon long before General Relativity and the  Robertson-Walker
metric \cite{Calinon}.

   The group theory viewpoint came into its own with
Einstein's theory of Special Relativity, for which
$G= E(3,1)=ISO(3,1)$, the Poincar\'e group. Indeed only retrospectively
was the Galilei group recognised as its Wigner-\turk \cite{wigino}
contraction.
As with the Euclidean group,  the Poincar\'e group
admits two  continuous deformations, to $SO(4,1)$ or $ SO(3,2)$,
for which spacetime translations fail to commute.
It was perhaps only the early death  of Minkowski which
delayed until after the advent of Einstein's General
Relativity the implementation of Calinon's idea.
de Sitter, seeking a covariant version of Einstein's
Static Universe, introduced their cosets, de Sitter and anti-de Sitter
respectively.

   Einstein did not scruple to break boost invariance
with his static universe \cite{Einstein}, and this is a feature of
all Robertson-Walker metrics except those of de Sitter
\cite{deSitter}. A natural question, answered by Bacry and
Levy-Leblond \cite{Bacry:1968zf}, is what other 10-dimensional {\it
kinematical} algebras exist, that contain rotations, translations in
time and space  and boosts. All can be regarded as Wigner-\turk
contractions of the de Sitter and anti-de Sitter algebras.

Invariance under the local Lorentz group is extremely well attested
by experiment, but nevertheless Cohen and Glashow \cite{Cohen:2006ky}
observed that if it is broken down to its four-dimensional maximal
subgroup Sim$(2) \subset SO(3,1)$ it leaves invariant no spurion
fields, merely leaving fixed a null direction $n^\nu \equiv
\lambda n^\nu\,,\quad \lambda \ne 0$, where $\eta_{\mu \nu} n^\mu
n^\nu =0$. It may play a role linking small neutrino masses, and is
compatible with all present day tests of violations of Lorentz
invariance. Thus they proposed in their Very Special Relativity
theory that the fundamental local symmetry group is the semi-direct
product of Sim(2) and the translations, known as ISim$(2) \subset
ISO(3,1)$. In recent work, in an attempt to obtain non-commuting
translations, and hence spacetime curvature \cite{gibgompop}, we
studied the continuous deformations of ISim$(2)$ and found a
two-parameter family, one of which was rejected because the
deformation of the $SO(2)$  rotation generator ceased to be compact.
The remaining one-parameter deformed group DISim(2)$_b$ depends on a
dimensionless parameter $b$, and coincides with one introduced by
Bogoslovsky \cite{Bogoslovsky:2007gt} in his proposal for an
anisotropic Finslerian spacetime.

It is straight forward to generalise the ${\rm DIsim}(2)_b  $
group   to $k+2$ spacetime dimensions. We denote the resulting  group
$ {\rm DISim}_b(k)$. It is  interesting to note \cite{Kamimura}
that the  $ {\rm DISim}_b(k)$ is then isomorphic
to the extended Schr\"odinger group $ \tilde {\rm Sch}(k)$ \cite{Burdet}
which has resurfaced  in recent
studies of non-relativistic holography in $k$ spatial dimensions
(see e.g.  \cite{balasubra} and references therein).
Since this topic is not strictly connected with
the Maxwell algebra which is the main concern of the present paper, we
relegate  the details to appendix B.

   Since Maxwell's equations are invariant under DISim(2)$_b$, the dispersion
relation for photons is the standard one, and hence these theories are
consistent with the recent high-precision test of Lorentz violation
using the gamma-ray burst GRB090510 \cite{abdo}.

   The advent of quantum mechanics led to the realisation that
not only are deformations of algebras important, but so also
are  {\it extensions},
especially {\it central extensions}. The Ur-example  is
the Heisenberg algebra
\ben
[ \hat q^j , \hat p_i ] = i \hbar \delta^j_i \,.
\een
However, a more relevant example for our purposes is the motion of a
particle of charge $e$ in a uniform time-independent magnetic field.
The minimally-coupled Lagrangian is
\be\label{em0}
L= T+ e A_i\, \dot x^i\,,
\ee
where $A_i= -\ft12 F_{ij}\, x^j$, and $T$ is the kinetic energy.  If
$p_i\equiv \del T/\del \dot x^i$ (sometimes called the {\it mechanical
momentum}), then the {\it canonical momentum} is
\be
\pi_i\equiv \fft{\del L}{\del \dot x^i} =
p_i  + e A_i= p_i-\ft12 e F_{ij}\, x^j\,.
\ee
Assuming that $\pi_i$ and $x^j$ satisfy the standard Poisson algebra
$\{x^i,\pi_j\}=\delta^i_j$, $\{x^i,x^j\} = \{\pi_i,\pi_j\}=0$,
then $p_i$ and $x^j$ satisfy the centrally-extended algebra
\be
\{x^i,p_j\}=\delta^i_j\,,\qquad
\{x^i,x^j\}=0\,,\qquad
\{p_i, p_j\}= e F_{ij}\,.\label{heis}
\ee

   The action associated to (\ref{em0}) is invariant, up to a boundary term,
under constant translations:
\be
x^i\longrightarrow x^i + a^i\,.
\ee
The Noether charges associated with these spatial translations are
\be
\tilde p_i = \pi_i-\ft12 e F_{ij}\, x^j = p_i- e F_{ij}\, x^j\,.
\ee
It follows from the equations of motion that $d\tilde p_i/dt=0$.  One finds the
following non-trivial Poisson brackets:
\be
\{\tilde p_i, \tilde p_j\}= -e F_{ij}\,,\qquad \{x^i,\tilde p_j\}=\delta^i_j\,.
\label{poiss2}
\ee
Note that the $\tilde p_i$ and $p_i$ momenta Poisson commute:
\be
\{\tilde p_i,p_j\}=0\,.
\ee
The constancy of the $\tilde p_i$ is now seen to follow from the fact that
the Hamiltonian obtained by taking the Legendre transform of the Lagrangian
(\ref{em0}) is a function only of the mechanical momentum $p_i$,
and independent of the spatial coordinates $x^i$.

   One can think of $e$ as a central element $-\bar Z$ in a finite-dimensional
Poisson algebra, which commutes with all other generators.  The moment maps
$p_i$ and $\bar Z$ generate a group of transformations on phase space,
whose Lie algebra is
\be
[\bP_i,\bP_j] = \bZ\, F_{ij}\,, \qquad [\bP_i,\bZ]=0\,.\label{heis333}
\ee
(The relative sign between Lie algebra brackets and Poisson brackets is
a consequence of our general conventions, which are detailed in appendix A.)
Note that whereas the ``central term'' in the Poisson algebra
(\ref{heis333}) is cohomologically trivial (it can be removed by the local
redefinition of generators that maps from $x^i$ and $p_i$ to $x^i$ and
$\pi_i$), the central term in the Lie algebra (\ref{heis333}) is
cohomologically non-trivial.  The reason for the difference is that the
$x^i$ are included as generators in the Poisson algebra, but not in the
Lie algebra.

   Inclusion of a uniform time-independent electric as well as a
magnetic   field generalises (\ref{heis333})
to the Lorentz-covariant Lie algebra
\ben
\Bigl [ \bP _\mu , \bP _\nu \Bigr ] =  \bZ F_{\mu \nu} \,.
\label{elec}
\een
This five-dimensional algebra has as group manifold
$G$ the five-dimensional space with coordinates $x^\mu$ and $\theta$,
conjugate to $\bP_\mu$ and $\bZ$ respectively. It turns out, in
Kaluza-Klein fashion, that geodesic motion on $G$  projects down
onto  the coset $G/H$, where $H\equiv {\Bbb R}$ is generated by the
central element $ \bZ$, and electric charge is the conserved momentum
in that direction.
Particles with magnetic as well as electric charge (dyons) may also be
catered for, by passing to six
dimensions and replacing (\ref{elec}) by
\ben
\Bigl [ \bP _\mu , \bP _\nu \Bigr ] =
\bZ F_{\mu \nu} + \bZ ^\star F^\star_{\mu \nu}\,,
\label{elecmag}
\een
where $ F^\star_{\mu \nu} = \half \epsilon_{\mu \nu \alpha \beta }
F^{\alpha \beta}$ is the Hodge dual of $F_{\mu \nu}$.

   A different approach is to consider {\it non-central extensions}
of the fundamental algebra $\frak{g}$. This is by now standard
in supersymmetric $p$-brane  theories, following the pioneering work
of van Proeyen and van Holten \cite{vanPvanH}.
However the simplest example, which
is purely  bosonic and predates their work, is
the 16-dimensional Maxwell algebra, which is a non-central extension
of the Poincar\'e algebra with six tensorial charges $\bZ_{\mu\nu}$
arising through a non-commutativity of the momentum generators,
\be
{[} \bP_\mu, \bP_\nu {]} = \bZ_{\mu\nu}\,.
\ee
One now introduces six angles conjugate to $\bZ_{\mu \nu}$.
These angles
are dynamical variables with a non-trivial evolution.

 Note that for
any particular solution of the relevant equations of motion,
$\bar Z_{\mu\nu}=-e F_{\mu \nu}$, {\it spontaneous symmetry breaking} will
occur and the symmetry will be reduced to the subgroup of the
Poincar\'e group leaving the background $F_{\mu \nu}$ invariant.
This is the {\it kinematical group} in this context.

The aim of this paper is to study in the framework of the Very
Special Relativity \cite{Cohen:2006ky}, or its deformation
\cite{gibgompop}, the motion of a bosonic charged massive particle
in the presence of a constant electromagnetic field. This will
done by constructing the non-central extensions and deformations of
ISim(2). As we shall  see, the Maxwell-Sim algebra is constructed from
the translation generators $\bP_\mu$  and  the non-central extension
$ \bZ_{\mu\nu}=  [\bP_\mu,\bP_\nu]$, together with the Sim$(n)$
generators $(\bM_{+i}, \bM_{+-},\bM_{ij})$.


Later we study the deformations of the Maxwell-Sim algebra. In general
dimensions we find two deformations, with parameters $b$ and $c$. The
deformation parametrized by $c$ is analogous to the $k$ deformation of
the Maxwell algebra found in \cite{gomkamluk} (now restricted to the 14
generators of Maxwell-Sim(2)), which gave
$SO(3,2)\times SO(3,1)$ or  $SO(4,1)\times SO(3,1)$, depending on
the sign of $k$.
The $b$-deformation of Maxwell-Sim produces the Maxwell extension of
the DIsim$_b$ algebra, which is related to Finslerian geometry.

   In order to construct the particle models with the  previous symmetries,
we use two different approaches:  one based in the Lagrangian
formalism and  the non-linear realization approach \cite{Coleman},
and the other based on the  Hamiltonian formalism constructed from the
momentum maps.

In the case of Maxwell-DISim$_b$ the motion is given by a
Finslerian Lorentz force, while  for the undeformed Maxwell-Sim we obtain the
ordinary Lorentz force. Therefore the study of anisotropies of a
massive particle in an electromagnetic field could provide  a test of a
possible Finsler geometry.

The organization of the paper is as follows. In section 2 we review
the Maxwell algebra, and in section 3 we recall the basic facts about the
ISim algebra. In section 4 we construct the Maxwell-Sim algebra, and then in
section 5 we the study its deformations. The particle Lagrangians are
constructed in section 6, and in section 7 we perform  the Hamiltonian
analysis. The paper ends with conclusions.
There is also
an appendix about the Hamiltonian formalism and momentum maps.

\section{The Maxwell Algebra}

The name Maxwell algebra appears to originate with Glashow, as
reported in \cite{Stein} in connection with the behaviour of matter
in extremely  strong magnetic fields  such as are found in neutron
stars. Nowadays one might think of magnetars.
 However it is  Schrader
\cite{Schrader} who seems to have been the first to study it
systematically. Other earlier work applying group theoretic methods
to uniform electromagnetic fields is in \cite{Bacry,Bacry2}.
This often-cited work assumes a constant c-number background
field $F_{\mu \nu}$,  and is largely concerned with what they called
the {\it kinematical group}, i.e. with the
6-dimensional subgroup of the Poincar\'e group that leaves $F_{\mu \nu}$
invariant, generated by $\bP^\mu$ and two commuting Lorentz
generators
\be
\bG = \half F^{\mu \nu}\, \bM _{\mu \nu} \,,\qquad
\bstarG=
\half {\starF}^{\mu \nu} \,\bM_{\mu \nu} \,.
\ee
This gives the algebra
\bea
\bigl [\bG,\bstarG \bigr]  &=&0\,\\
\bigl [ \bG, \bP _\mu] &=& F_{\mu \nu} \bP ^\nu\\
\bigl [ \bstarG, \bP _\mu] &=& {\starF}_{\mu \nu} \bP ^\nu\,,
\eea
where we have defined $\starF_{\mu\nu}= \ft12 \epsilon_{\mu\nu\rho\sigma}\,
F^{\rho\sigma}$.
We shall refer to these  6-dimensional kinematical algebras
 as the Bacry-Combe-Richards or BCR-algebras.
The  BCR group is $E(2) \times E(1,1)$,
the product of the two-dimensional  Euclidean group
$E(2)$ with the two-dimensional Poincar\' e group $E(1,1)$.

   If we use light-cone coordinates where $x^\mu=(x^-,x^+,x^i)$, define
$\epsilon_{+-12}=+1$, and
take the Maxwell field to be zero except for $F^{+-}=1$, then
$\bG=\bM_{+-}$, $\bstarG=\bM_{12}$, and the algebras associated with
the two factors are generated by
\bea
E(2):&& \{\bP_1,\bP_2,\bstarG\}\,,\nn\\
E(1,1):&& \{\bP_+,\bP_-,\bG\}\,.
\eea
For more details on the BCR group, the reader may consult \cite{Janner}.

According to \cite{Combe}, the BCR algebra has
three central extensions, so that

\be
\bigl [\bG,\bstarG \bigr]  = {\bf a} \,,\qquad
\bigl [\bP_\mu , \bP _\nu] =  {\bf Z_{elec}}  F_{\mu \nu} +
              {\bf Z_{mag}} {\starF}_{\mu \nu}
 \,.
\ee
There also four other non-central charges ${\bf Z_{\mu\nu}}$ (in the complement
of ${\bf Z_{\mu\nu}^{elec}} = {\bf Z_{elec}} F_{\mu \nu}$ and
 ${\bf Z_{\mu\nu}^{mag}} = {\bf Z_{mag}} {\bf{\starF}}_{\mu \nu}$):
\bea
 \bigl [\bP_\mu , \bP _\nu] &=& {\bf Z_{elec}} F_{\mu \nu} +
              {\bf Z_{mag}} {\starF}_{\mu \nu}+ {\bf Z_{\mu\nu}}
 \,.
\eea
In total there are six extensions which we can decompose in
representations of $\bG$ and $\bstarG$. The difference with respect to the
Maxwell case to be treated later is the presence of two central charges.
The central
charges are present because  the Lorentz group has been reduced to
the abelian subgroup $G_2$ generated  by  $\bG$ and $\bstarG$. (Note that
this 2-dimensional abelian group $G_2$ is not to be confused with the
14-dimensional non-abelian simple Lie group of the Cartan/Dynkin
classification!)

In \cite{Combe,Hoogland}  arguments are given
to the effect  that if the rotation in $\bG
,\bstarG $ is to be a compact generator then ${\bf a}$ should
vanish. That leaves ${\bf Z_{elec}}$  and ${\bf Z_{mag}}$ which,
as the notation suggests,  may be identified
with electric and magnetic charge respectively.
In what follows we shall refer to the  8-dimensional
doubly-extended  kinematic algebra
generated by $\{\bP_\mu, \bG, \bstarG,{\bf Z_{elec}},{\bf Z_{mag}}\}$ as
the EBCR algebra.

  The EBCR algebra is a direct sum of two subalgebras, each of which has
a non-trivial quadratic Casimir.  In the case where the only non-vanishing
component of $F^{\mu\nu}$ is given by $F^{+-}=1$, the generators of the
two subalgebras are $\{\bstarG, \bP_1,\bP_2,{\bf Z_{mag}}\}$ and
$\{\bG,\bP_+,\bP_-,{\bf Z_{elec}}\}$.  The two Casimirs are
\be
\bC_{\rm mag} = \ft12 \bP_i^2 + {\bf Z_{mag}}\, \bM_{12}\,,\qquad
\bC_{\rm elec} = \bP_+\, \bP_-  - {\bf Z_{elec}}\, \bM_{+-}\,.
\ee

   By contrast, the  Maxwell algebra  is a 16-dimensional extension of the
Poincar\'e algebra  with the non-vanishing brackets
\bea
\bigl [ \bM_{\alpha \beta} ,\bP_\gamma \bigr ]&=&
\eta_{\alpha \gamma} \bP_\beta -\eta_{\beta \gamma} \bP _\alpha\,,\\
\bigl [ \bM_ {\alpha \beta}, \bM_{\gamma \delta} \bigr ] &=&
\eta_{\alpha \gamma} \bM_{\beta \delta} - \eta _{\alpha \delta}
\bM_{\beta \gamma} + \eta_{\beta \delta} \bM_{\alpha \gamma  }  - \eta_{\beta
\gamma} \bM_{\alpha \delta}\,,\\
\bigl [ \bM_ {\alpha \beta}, \bZ_{\gamma \delta} \bigr ] &=&
\eta_{\alpha \gamma} \bZ_{\beta \delta} - \eta _{\alpha \delta}
\bZ_{\beta \gamma} + \eta_{\beta \delta} \bZ_{\alpha \gamma  }  - \eta_{\beta
\gamma} \bZ_{\alpha \delta}\,,\\
\bigl [\bP_\alpha, \bP_\beta \bigr ] &=& \bZ_{\alpha \beta} \,,
\eea
where  $\bZ_{\alpha\beta}=- \bZ_{\beta\alpha}$.
There are two generic Casimirs (which exist for the Maxwell algebra in
any dimension)  \cite{Schrader},
\ben
\bC_1 =  \ft12 \bigl ( \bP^\mu \bP_\mu + \bM_{\mu \nu} \bZ^{\mu \nu} \bigr )
\,,\qquad
  \bC_2= \half \bZ_{\mu \nu} \bZ^{\mu \nu} \,,
\een
and a third quadratic Casimir that exists only in the special case of four dimensions:
\be
\bC_3=\ft12 \bZ_{\mu\nu} \bstarZ^{\mu\nu}\,.
\ee
The Casimirs are, of course, elements of the universal
enveloping algebra of the Maxwell Lie algebra.

The six additional generators $\bZ_{\mu \nu}$ are on the
same footing as the Poincar\'e generators
$\bP_\mu$ and $\bM_{\mu\nu}$, in that they are
dynamical; depending upon the equations of motion of the theory under
consideration, the associated momenta may vary with time.
The  $\bZ_{\mu \nu}$ are associated with the Maxwell
2-form. In specific solutions of any equations of motion, there may
occur spontaneous symmetry breaking in which the associated momenta
$\bar Z_{\mu\nu}$ take  constant
values, $\bar Z_{\mu \nu}= -e F_{\mu \nu}$.\footnote{More precisely, in a
Hamiltonian treatment the $\bar Z_{\mu\nu}$ are the moment maps.
See section \ref{hamilsec} for further details.} The relevant algebra will then
reduce to the EBCR algebra that leaves invariant the background
field $F_{\mu \nu}$.

  A set of left-invariant 1-forms are (we omit those for the
Lorentz sub-algebra)
\bea\label{mcformsmaxwell}
P ^\mu &=& d x ^\mu \\
Z ^{\mu \nu} &=&d \theta ^{\mu \nu} - \half (x^\mu d x^\nu -x^\mu d x ^\mu) \,,
\eea
with generators of right actions being given by
\bea
P_{\mu}&=&{\p \over \p x ^\mu } - \half  x^\nu
{\p  \over \p \theta ^{\mu \nu} }\\
Z_{\mu \nu} &= & { \p \over \p \theta ^ {\mu \nu} } \,.
\eea
they satisfy $ \bigl [P_\mu,P_\nu \bigr ]=Z_{\mu \nu}$.

    The 1-forms (\ref{mcformsmaxwell}) are
invariant under
\bea
\delta x^\mu &=&\epsilon^\mu\,,\nn\\
 \delta
\theta^{\mu\nu}&=&\epsilon^{\mu\nu}
+\half (\epsilon^\mu x^\nu-\epsilon^\nu x^\mu)\,,
\eea
which are generated by the vector fields
\bea
P_{\mu}&=&{\p \over \p x^\mu } + \half  x^\nu
{\p  \over \p \theta ^{\mu \nu} }\,,\nn\\
Z_{\mu \nu} &= & { \p \over \p \theta ^ {\mu \nu} } \,.
\eea

 The ten-dimensional subalgebra, which is obtained by taking the
quotient with respect to the Lorentz subalgebra,  is spanned by
$\bP^\mu$ and $\bZ^{\mu \nu}$, and closes on the generalised Heisenberg
algebra. The associated  coset is thus also a group manifold,
sometimes called a {\sl superspace}, and has as coordinates $x^\mu$
and $\theta ^{\mu \nu}$. This 10-dimensional superspace, which is
fibred over Minkowski spacetime with flat six-dimensional fibres,
carries a natural Lorentz-invariant metric:
\ben
ds ^2_{10} = \eta_{\mu \nu} P^\mu P^\nu + \half Z^{\mu \nu} Z_{\mu  \nu}
\,.
\een

\subsection{Quantisation}

The obvious approach to quantisation is to
consider wave functions $\Psi(x^\mu, \theta ^{\mu \nu})$  depending upon both
$x^\mu$ and $\theta ^{\mu \nu}$. A generalised Klein-Gordan or Dirac
equation  may readily  be written in the usual way using the
differential  operators
\ben
P_\mu\,, \qquad Z_{\mu \nu} \,.
\een
The equations can be solved using Fourier transforms,
and the solutions used to construct one-particle Hilbert spaces.
The Maxwell group acts on these wave functions by
pull-back, and in this way one obtains a {\sl projective}
representation of the Maxwell group. For details
of the procedure, including the calculation of the relevant
co-cycles, the reader is referred to Schrader's paper \cite{Schrader}.

\subsection{Deformations and Contractions}

In general dimensions, the Maxwell algebra admits a unique
deformation parameter $k$.  For $k>0$ we have
 $\frak{so}  (D-1,2) \oplus\frak{so}( D-1,1) = ({\cal
M}_{\mu \nu}, {\cal P}_\mu ; {\cal J} _{\mu \nu} )$.  If instead $k<0$,
we have
$\frak{so}( D,1) \oplus\frak{so}( D-1,1)$,  where $D=n+2$ is the dimension of space time,
\cite{Soroka:2006aj,gomkamluk}. Conversely, it may be regarded as a
Wigner-\turk contraction \cite{gomkamluk} such that
\bea
M_{\mu \nu} &=& {\cal M}_{\mu \nu} \pm {\cal J}_{\mu \nu} \,, \\
 P_\mu &=& \lim _{k \rightarrow\infty} { {\cal P}_\mu  \over |k|}\\
Z_{\mu \nu} &=& \lim_ {k \rightarrow \infty}- {{\cal M}_{\mu \nu} \over k^2 }
\eea
where $k$ has the dimensions of length and the sign choice is
made depending on whether we consider the
AdS or dS part.


\section{The ISim Algebra}

   We may consider a generalisation of the discussion of the Maxwell
algebra of \cite{Bonanos:2008kr}, where the starting point is
taken to be the ISim algebra rather than the Poincar\'e algebra.
The ISim generators are
\be
\bP_\mu\,,\qquad \bM_{+i}\,,\qquad \bM_{+-}\,,\qquad \bM_{ij}\,.
\ee
The ISim algebra, with the conventions we are using, is given in
\cite{gibgompop}.

We define left-invariant 1-forms $\lambda$ as in (\ref{lamrhodef}),
but now, for convenience, we denote them by $\lambda^a=(P^\mu,M^{+i},M^{+-},
  M^{ij})$, and so
\be
g^{-1} dg = P^\mu\, \bP_\mu + M^{+i}\, \bM_{+i} +
 \ft12 M^{ij}\, \bM_{ij}  + M^{+-}\, \bM_{+-}\,.
\ee
In terms of these, the ISim$(n)$ algebra is given by
\bea
P^+ &=& M^{+i}\wedge P^i +
         M^{+-}\wedge P^+\,,\qquad
dP^- = - M^{+-}\wedge P^-\,,\nn\\
dP^i &=& M^{ij}\wedge P^j
     - M^{+i}\wedge P^-\,,\nn\\
d M^{+i} &=& M^{ij}\wedge M^{+j}
      + M^{+-}\wedge M^{+i}\,,\nn\\
dM^{+-} &=&0\,,\qquad dM^{ij}=M^{ik}\wedge M^{kj}\,.\label{ISimn}
\eea

\section{The Maxwell-Sim Algebra}

   The Maxwell-Sim algebra can be constructed in complete analogy
to the Maxwell algebra discussed previously.  One way to describe this
it that we start with the $\bP_\mu$ generators alone, obtain the
central extension in which $[\bP_\mu,\bP_\nu]= \bZ_{\mu\nu}$, and
then append the Sim$(n)$ generators $(\bM_{+i}, \bM_{+-},\bM_{ij})$ to
form the Maxwell-Sim$(n)$ algebra.  At the level of the left-invariant
1-forms, this means that we augment the ISim$(n)$ relations (\ref{ISimn})
by
\bea
dZ^{+i}&=& -P^+\wedge P^i + M^{+-}\wedge Z^{+i} + M^{ij}\wedge Z^{+j} +
             M^{+i}\wedge Z_{-+} -  M^{+j}\wedge Z^{ij}\,,\nn\\
dZ^{-i} &=& -P^-\wedge P^i - M^{+-}\wedge Z^{-i} + M^{ij}\wedge M^{+j}\,,\nn\\
dZ^{-+} &=& - P^-\wedge P^+ + M^{+i}\wedge Z^{-i}\,,\nn\\
dZ^{ij} &=& - P^i\wedge P^j - M^{+i}\wedge Z^{-j} + M^{+j}\wedge Z^{-i}\,.
\label{dZ}
\eea

\section{Deformations of Maxwell-Sim$(n)$}

   We follow the method for finding the general non-trivial deformation
of an algebra that is described in \cite{gibgompop} (see \cite{levnah} for
further details).  This entails
first finding the second cohomology class $H^2(\frak{g},\frak{g})$, which
determines the non-trivial deformations at the linear level.  If
$H^3(\frak{g},\frak{g})$ is trivial,
then there must exist, possibly after making
(trivial) redefinitions, an extension of the linearised deformations
that is valid to all orders.  (This is checked by verifying that the
deformed algebra satisfies the Jacobi identities.)  If, on the other
hand, $H^3(\frak{g},\frak{g})$ is non-trivial, then the extension
beyond the linearised level may not be possible.

   For the generic case of the Maxwell-Sim$(n)$ algebra, we find
that there are two distinct 1-parameter non-trivial deformations. We
denote these by the $b$-deformation and the $c$-deformation, where
$b$ and $c$ are the respective constants parameterising the two
deformations.\footnote{We have performed some of the calculations
with differential
forms with the aid of the EDC Mathematica package \cite{Bonanos}.}

\subsection{The $b$-deformation}

   In the $b$-deformation, the Maxwell-Sim$(n)$ algebra defined by
(\ref{ISimn}) and (\ref{dZ}) is modified by the following additions
to $dP^\mu$ and $dZ^{\mu\nu}$:
\bea
dP^\mu &=& b M^{+-}\wedge P^\mu +\cdots\,,\nn\\
dZ^{\mu\nu} &=& 2b M^{+-}\wedge Z^{\mu\nu} +\cdots\,,\label{bdeform}
\eea
where the ``$\cdots$''
terms represent the usual right-hand sides of the undeformed
Maxwell-Sim$(n)$ algebra.  The Sim$(n)$ relations in (\ref{ISimn}) are
unmodified.

\subsection{The $c$-deformation}

  In the $c$-deformation, the Maxwell-Sim$(n)$ algebra defined by
(\ref{ISimn}) and (\ref{dZ}) is modified by the following additions
to $dP^\mu$ and $dZ^{\mu\nu}$:
\bea
dP^\mu &=& c P_\nu\wedge Z^{\mu\nu} +\cdots\,,\nn\\
dZ^{\mu\nu} &=& -c Z^\mu{}_\rho \wedge Z^{\rho\nu} +\cdots\,,\label{cdeform}
\eea
where again the ``$\cdots$'' terms represent the usual right-hand sides of
the undeformed Maxwell-Sim$(n)$ algebra.
The Sim$(n)$ relations in (\ref{ISimn}) are again unmodified.


   It is not possible to turn on the $b$ and $c$ deformations simultaneously.
(This agrees with the fact that there are no
de Sitter or anti-de Sitter deformations of the ISim algebra \cite{gibgompop}.)

   In the special case of Maxwell-Sim(2), we find that there is an
additional non-trivial deformation characterised by a parameter $a$, which
can be turned on simultaneously with the $b$-deformation.  Thus in place
of the $b$-deformation given by (\ref{bdeform}), for Maxwell-Sim(2) we may
have
\bea
dP^\mu &=& a M^{12}\wedge P^\mu + b M^{+-}\wedge P^\mu +
       \cdots\,,\nn\\
dZ^{\mu\nu} &=& 2 a M^{12}\wedge Z^{\mu\nu}+ 2 b M^{+-}\wedge Z^{\mu\nu}
 + \cdots\,.\label{abdeform}
\eea

   A calculation of the cohomology group $H^3(\frak{g},\frak{g})$
for Maxwell-Sim(2) shows that it is non-trivial, and of dimension 3.

  Note that the deformation parameterised by $c$ in (\ref{cdeform})
is analogous to the $k$ deformation of the Maxwell algebra
found in \cite{gomkamluk}, now restricted to the 14 generators of
Maxwell-Sim(2), which gave $SO(3,2)\times SO(3,1)$ or
$SO(4,1)\times SO(3,1)$ depending on the sign of $k$.

  The deformations associated with the parameters $a$ and $b$ are
the Maxwellian extensions of the $a$ and $b$ deformations
of the ISim(2) algebra obtained in \cite{gibgompop}.

\section{Lagrangians}

\subsection{The Maxwell case}

 A particle  model can be derived geometrically
 by the techniques of non-linear realisations,
 \cite{Coleman}.
Let us first consider the coset (Maxwell)/(Lorentz). In order
to construct a Lorentz invariant Lagrangian from
the Maurer-Cartan forms (\ref{mcformsmaxwell}), one possibility is
to introduce new dynamical variables $f_{\mu\nu}$ that transform covariantly
under the Maxwell group \cite{bongom2}.
The Lagrangian becomes
\be
L=-m\sqrt{-\dot
x^2}+\frac12f_{\mu\nu}\left(\dot\theta^{\mu\nu}-\frac12(x^\mu\dot
x^\nu- x^\nu\dot x^\mu)\right). \label{Lag}
\ee
The equations of motion in the proper time gauge are
\bea
\label{eqmotion1theta} \dot f_{\mu\nu}&=&0,
\\\label{eqmotion1f} \dot
\theta^{\mu\nu}&=&\frac12(x^\mu\dot x^\nu-x^\nu\dot x^\mu),
\\
\label{eqmotion1x} m\ddot x_\mu&=&f_{\mu\nu}\dot x^\nu\,. \label{eqmotion1xx}
\eea
 Integration of (\ref{eqmotion1theta}) gives
$f_{\mu\nu}=f^0_{\mu\nu}$, and such a solution spontaneously breaks the Lorentz
symmetry into a subalgebra of the Maxwell algebra (namely the EBCR
algebra discussed earlier). Substituting this solution into
equation
(\ref{eqmotion1xx}) gives  the motion of a particle in a
constant electromagnetic field.

Alternatively, since we know how to construct Lorentz scalars, we can
construct a Lagrangian without the introduction of the new dynamical
variables $f_{\mu\nu}$ as
\be L=m\dot
x^2+\frac{\alpha}{2}{\left(\dot\theta^{\mu\nu}-\frac12(x^\mu\dot
x^\nu- x^\nu\dot x^\mu)\right)}^2\,. \label{Lag0}
\ee
The quantities
 $\dot\theta^{\mu\nu}+\frac12(x^\mu\dot x^\nu- x^\nu\dot x^\mu)$ are
 constants of motion.  If we choose them equal to $\frac 12 f^0_{\mu\nu}$,
  we recover the same equation of motion of a particle moving in a
 constant electromagnetic field that we obtained above.

  Another way to construct the Lagrangian is to consider the coset
(Maxwell)/(Rotations). This coset is useful for the construction of massive
particle Lagrangians when the tensor calculus is not known.
(For example, in the case of ISim, we may consider
the coset (ISim)/(Rotations), rather than (ISim)/(Sim), because we do not know
{\it a priori} what is the
length element; in order words, we do not have an obvious tensorial
calculus. A more striking example is the case of the deformed ISim
algebra DISim$_b$, discussed in \cite{gibgompop}.
We obtain left-invariant 1-forms, by first defining
\be
g= g_0\, U\,,
\ee
where
\be
g_0= e^{x^\mu P_\mu}\, e^{\ft12\theta^{\mu\nu}
Z_{\mu\nu}}\,,\qquad U=e^{w^i M_{0i}}\,,
\ee
and $w^i$, $i=1,2,3$, are the Goldstone bosons associated with the broken boost
generators.  The left-invariant 1-forms $\lambda$ may then be read off
from
\bea
g^{-1} dg &=& U^{-1}\, g_0^{-1}\, dg_0\, U +
    U^{-1}\, dU\,,
\nn\\
&=&\lambda^\mu_P\, P_\mu + \lambda^{\mu\nu}_Z\, Z_{\mu\nu}
+\lambda^i_M\, M_{0i}  + \lambda^{ij}_R\, M_{ij} \,.
\eea
Defining
\be
 g_0^{-1}\, dg_0 = \overline{\lambda}^\mu_P\, P_\mu
    + \overline{\lambda}^{\mu\nu}_Z\, Z_{\mu\nu}\,,
\ee
we have
\be
 \overline{\lambda}^\mu_P= dx^\mu\,,\qquad
 \overline{\lambda}^{\mu\nu}_Z = d\theta^{\mu\nu} -
       \ft12 (x^\mu\, dx^\nu - x^\nu\, dx^\mu)\,.
\ee
The Lorentz transformations generated by $U$ may be used to define
$\Lambda^\nu{}_\mu(w^i)$:
\be
U^{-1}\, P_\mu U = \Lambda^\nu{}_\mu(w^i)\, P_\nu\,.
\ee

The left-invariant 1-forms $\lambda^\mu_P, \lambda^{ij}_R\,$ and $ \lambda^{\mu\nu}_Z$
are then given by
\bea \lambda^\mu_P &=& \Lambda^\mu{}_\nu\, \overline{\lambda}^\nu_P
=
         \Lambda^\mu{}_\nu\, dx^\nu\,,\nn\\
\lambda^{\mu\nu}_Z &=& \Lambda^\mu{}_\rho\, \Lambda^\nu{}_\sigma\,
 \overline{\lambda}^{\rho\sigma}_Z =
  \Lambda^\mu{}_\rho\, \Lambda^\nu{}_\sigma\,
  [ d\theta^{\rho\sigma} -
       \ft12 (x^\rho\, dx^\sigma - x^\sigma\, dx^\rho)]\,.
\eea
The 1-forms $\lambda^i_M, $ are given by

\bea  \lambda^i_M&=&  dw^i\,+ dw^j (\delta^i_j\,-w^i\,w^j)(\frac{sinh\,w}{w}-1) \nn\\
\lambda^{ij}_R&=& (\frac{dw^i\,w^j-dw^j\,w^i}{w^2})(cosh w-1)
\eea

A particle Lagrangian that is invariant under $SO(3)$  is
\be L = \Big[\alpha\, \lambda^0_P+
      \ft12 \hat f_{\mu\nu}(\tau)\, \lambda^{\mu\nu}_Z\Big]^*\,,\label{Lag1}
\ee
where the $*$ indicates that the 1-forms are pulled back onto the
world-line: $[dx^\mu]^* \equiv \dot x^\mu(\tau)\, d\tau$, etc.  The
coefficient $\alpha$ is constant, whilst $\hat f_{\mu\nu}(\tau)$ is
a dynamical field that depends upon $\tau$. We see that (\ref{Lag1})
may be written as
\be L= \alpha\, \Lambda^0{}_\mu\, \dot x^\mu  + \ft12 f_{\mu\nu}\,
 [\dot\theta^{\mu\nu} -\ft12 (x^\mu\, \dot x^\nu -x^\nu\, \dot x^\mu)]\,,
\ee
where $\Lambda^\mu{}_\nu$ is a general Lorentz boost transformation
and depends on the non-dynamical coordinates $w^i$.  We have also
introduced the tensor field $f_{\mu\nu}$, which is related to $\hat
f_{\mu\nu}$ by
\be f_{\mu\nu}= \Lambda^\rho{}_\mu(w^i)\,
\Lambda^\sigma{}_\nu(w^i)\,
                      \hat f_{\rho\sigma}\,.
\ee

   We now define the particle momentum $p_\mu$ in the canonical way:
\be p_\mu= \fft{\del L}{\del \dot x^\mu} = \alpha\, \Lambda^0{}_\mu
          + {\bf \ft12} f_{\mu\nu}\, x^\nu\,.
\ee
Because $\Lambda^0{}_\mu$ is a timelike  Lorentz
vector, we have
\be
(p_\mu - \ft12 f_{\mu\nu}\, x^\nu)^2 = -m^2\,.\label{mass}
\ee
Introducing $e$ as a Lagrange multiplier to enforce the mass-shell
condition (\ref{mass}), we arrive at the Lagrangian
\crampest
\be L= (p_\mu -  \ft12 f_{\mu\nu}\, x^\nu)\, \dot x^\mu  + \ft12
f_{\mu\nu}\,
 [\dot\theta^{\mu\nu} -\ft12 (x^\mu\, \dot x^\nu -x^\nu\, \dot x^\mu)]
-\ft12 e\, [(p_\mu -  \ft12 f_{\mu\nu}\, x^\nu)^2 + m^2]\,.\label{Lag2}
\ee
\uncramp

  Varying with respect to $p_\mu$ gives
\be \dot x^\mu = e(p^\mu - \ft12 f^\mu{}_\nu\, x^\nu)\,. \ee
Substituting for $p_\mu$ in (\ref{Lag2}), and then varying with
respect to $e$ to obtain
\be
e= -\fft{\sqrt{-\dot x^2}}{m}\,,
\ee
we finally arrive at the Lagrangian (\ref{Lag}). In section 7  we
shall see how the non-linear realisation method and coadjoint orbit
technique gives the same results.


\subsection{The Maxwell-Sim Lagrangian}

 We  start with the coset (Maxwell-Sim)/$SO(2)$, and
then construct the left-invariant 1-forms from the coset representative
\be
g= g_0\, U\,,
\ee
with
\be
g_0= e^{x^\mu P_\mu}\, e^{\ft12\theta^{\mu\nu}
Z_{\mu\nu}}\,,\qquad U=e^{w^i M_{+i}}\, e^{w N}\,.
\ee
Following the same steps as in the Maxwell case, we have
\bea
g^{-1} dg &=&\lambda^\mu_P\, P_\mu + \lambda^{\mu\nu}_Z\,
Z_{\mu\nu} +\lambda^i_M\, M_{+i} +
      \lambda_N\, N \,.
\eea
The left-invariant 1-forms $\lambda^\mu_P$ and $ \lambda^{\mu\nu}_Z$
are then given by
\bea
\lambda^\mu_P &=& \Lambda^\mu{}_\nu\, \overline{\lambda}^\nu_P
=
         \Lambda^\mu{}_\nu\, dx^\nu\,,\nn\\
\lambda^{\mu\nu}_Z &=& \Lambda^\mu{}_\rho\, \Lambda^\nu{}_\sigma\,
 \overline{\lambda}^{\rho\sigma}_Z =
  \Lambda^\mu{}_\rho\, \Lambda^\nu{}_\sigma\,
  [ d\theta^{\rho\sigma} -
       \ft12 (x^\rho\, dx^\sigma - x^\sigma\, dx^\rho)]\,,
\eea
where
\be
 \overline{\lambda}^\mu_P= dx^\mu\,,\qquad
 \overline{\lambda}^{\mu\nu}_Z = d\theta^{\mu\nu} -
       \ft12 (x^\mu\, dx^\nu - x^\nu\, dx^\mu)\,.
\ee
The Lorentz transformation $ \Lambda^\mu{}_\nu (w^i,w)$ is given by
\be \Lambda^\mu{}_\nu = \begin{pmatrix} e^{-w} \quad& 0\quad
                        & 0\\
        -\ft12e^w \,w^k w^k\quad & e^{w} \quad & e^w\, w^i\\
        -w^j\quad & 0\quad & 1 \end{pmatrix}\,,\label{Lammata}
\ee
where we order the spacetime coordinates in the sequence
$x^\mu=(x^-,x^+,x^i)$, $i=1,2$.

 The 1-forms $\lambda^i_M$ and $\lambda_N$ are given by
\be
\lambda^i_M= e^w\, dw^i\,,\qquad \lambda_N = dw\,.
\ee

A particle Lagrangian that is invariant under $SO(2)$ (generated by
$J=M_{12}$) is given by
\be
L = \Big[\alpha\, \lambda^+_P - \beta \lambda^-_P +
      \ft12 \hat f_{\mu\nu}(\tau)\, \lambda^{\mu\nu}_Z\Big]^*\,.\label{Lag1a}
\ee
As in the Maxwell case  the coefficients $\alpha$ and
$\beta$ are constants, whilst $\hat f_{\mu\nu}(\tau)$ is a dynamical
field that depends upon $\tau$. We see that (\ref{Lag1a}) may be
written as
\be
L= \alpha\, \Lambda^+{}_\mu\, \dot x^\mu -
     \beta\, \Lambda^-{}_\mu\, \dot x^\mu + \ft12 f_{\mu\nu}\,
 [\dot\theta^{\mu\nu} -\ft12 (x^\mu\, \dot x^\nu -x^\nu\, \dot x^\mu)]\,,
\ee
where
\be f_{\mu\nu}= \Lambda^\rho{}_\mu(w,w^i)\,
\Lambda^\sigma{}_\nu(w,w^i)\,
                      \hat f_{\rho\sigma}\,.
\ee

   The particle momentum $p_\mu$ is given by
\be
p_\mu= \fft{\del L}{\del \dot x^\mu} = \alpha\, \Lambda^+{}_\mu
          -\beta\, \Lambda^-{}_\mu +  \ft12 f_{\mu\nu}\, x^\nu\,.
\ee
Noting that $\Lambda^+{}\mu\, \Lambda^{+\mu}= \Lambda^-{}_\mu\,
  \Lambda^{-\mu}=0$ and $\Lambda^+{}_\mu\, \Lambda^{-\mu}=1$, we see that
\be
(p_\mu -  \ft12 f_{\mu\nu}\, x^\nu)^2 = -m^2\,,\label{mass1}
\ee
where we have defined the mass parameter as
\be
m=\sqrt{2\alpha\,\beta}\,.
\ee
Introducing $e$ as a Lagrange multiplier to enforce the mass-shell
condition (\ref{mass1}), we arrive at the Lagrangian
\crampest
\be
L= (p_\mu -  \ft12 f_{\mu\nu}\, x^\nu)\, \dot x^\mu  + \ft12
f_{\mu\nu}\,
 [\dot\theta^{\mu\nu} -\ft12 (x^\mu\, \dot x^\nu -x^\nu\, \dot x^\mu)]
-\ft12 e\, [(p_\mu -  \ft12 f_{\mu\nu}\, x^\nu)^2 + m^2]\,.\label{Lag2a}
\ee
\uncramp

  Varying with respect to $p_\mu$ gives
\be
\dot x^\mu = e(p^\mu -  \ft12 f^\mu{}_\nu\, x^\nu)\,.
\ee
Substituting for $p_\mu$ in (\ref{Lag2a}), and then varying with
respect to $e$ we get the Lagrangian (\ref{Lag}). Thus the
undeformed Maxwell-Sim algebra gives the same particle Lagrangian as the
Maxwell algebra based on the full Poincar\'e group.

%

\subsection{The Maxwell-DISim$_b$ Lagrangian}

    The left-invariant 1-forms $\lambda^\mu_P$  of the DISim$_b$ algebra
are given by
\bea
\lambda^\mu_P &=&
         \tilde\Lambda^\mu{}_\nu\, dx^\nu\,,
\eea
where the matrix $\tilde\Lambda$ is
\be
\tilde\Lambda^\mu{}_\nu = \begin{pmatrix} e^{-w(1-b)} \quad&
0\quad
                        & 0\\
        -\ft12e^{w(1+b)} \,w^k w^k\quad & e^{w(1+b)} \quad & e^{w(1+b)}\, w^i\\
        -w^j\quad & 0\quad & 1 \end{pmatrix}\,.\label{Lammat}
\ee

 The 1-forms $\lambda^i_M$
and $\lambda_N$ are given by
\be
\lambda^i_M= e^w\, dw^i\,,\qquad \lambda_N = dw\,.
\ee

   We wish to construct a particle Lagrangian that is invariant under
$SO(2)$ (generated by $J=M_{12}$). Thus we begin by writing
\be
L = \Big[\alpha\, \lambda^+_P - \beta \lambda^-_P
\Big]^*\,,\label{Lag1b}
\ee
where as before the $*$ indicates that the 1-forms are pulled back onto the
world-line.  The
coefficients $\alpha$ and $\beta$ are constants. We see that
(\ref{Lag1b}) may be written as
\be
L= \alpha\, \tilde\Lambda^+{}_\mu\, \dot x^\mu -
     \beta\, \tilde\Lambda^-{}_\mu\, \dot x^\mu \,,
\ee
where $\tilde\Lambda^\mu{}_\nu$ depends on the non-dynamical
coordinates $w$ and $w^i$, and is given by (\ref{Lammat}).

   We now define the particle momentum $p_\mu$ in the canonical way:
\be
p_\mu= \fft{\del L}{\del \dot x^\mu} = \alpha\,
\tilde\Lambda^+{}_\mu
          -\beta\, \tilde\Lambda^-{}_\mu \,.
\ee
Noting that $\tilde\Lambda^+{}\mu\, \tilde\Lambda^{+\mu}=
\tilde\Lambda^-{}_\mu\,
  \tilde\Lambda^{-\mu}=0$ and
\be
\tilde\Lambda^+{}_\mu\, \tilde\Lambda^{-\mu}= e^{2wb}\,,
\ee
we have the constraint
\be
p^2 = -2\alpha\beta{\Big(\frac{p_+}{\alpha}}\Big)^{\frac{2b}{1+b}}=
-2\alpha\beta{\Big(\frac{n^\mu p_\mu}{\alpha}}\Big)^{\frac{2b}{1+b}}\,.
\label{mass2}
\ee
With $\alpha=-m(1-b)$ and $\beta=-\ft12 m(1+b)$ we obtain equation
(18) of \cite{gibgompop}:
\be
\label{mass3}
p^2+m^2(1-b^2)\Big(-\fft{n^\nu p_\nu}{m(1-b)}
       \Big)^{2b/(1+b)}=0\,.
\ee
Introducing $e$ as a Lagrange multiplier to enforce the mass-shell
condition (\ref{mass3}), we arrive at the Lagrangian
\crampest
\be L= p_\mu\, \dot x^\mu -\ft12 e\, \Big[p^2+m^2(1-b^2)\Big(-\fft{n^\nu p_\nu}{m(1-b)}
       \Big)^{2b/(1+b)} \Big]\,.
\label{Lag2b}
\ee
\uncramp
Varying with respect to $p_\mu$ gives
\be
\dot x^\mu = e\Big[ p^\mu - b\, m\, \Big(-\fft{n^\nu p_\nu}{m(1-b)}
        \Big)^{\fft{b-1}{b+1}}\, n^\mu \Big]\,.
\label{vexp}
\ee
If we solve for $p_\mu$  and substitute into
(\ref{Lag2b}), we obtain
\crampest
\be L= \ft12 \frac{\dot{x}^2}{e} -
\ft12 m^2(1-b^2)\Big(-\fft{n^\nu \dot{x}_\nu}{m(1-b)}
       \Big)^{2b/(1+b)}\,e^{(1-b)/(1+b)}\,.
\label{Lag3b}
\ee
\uncramp
Varying this with respect to $e$ we get
\be
e= \fft1{m(1-b)}\, \Big(-\dot{x}^2
        \Big)^{\fft{1+b}{2}}\,\Big(-n^\nu \dot{x}_\nu
        \Big)^{-b}\,,
\label{eexp}
\ee
from which we obtain the Finslerian Lagrangian of \cite{gibgompop}
\be
L= -m  (-\eta_{\mu\nu} \dot x^\mu \dot x^\nu)^{(1-b)/2}\,
     (-n_\rho \dot x^\rho)^b\,.\label{Lag4b}
\ee

      For the Maxwell-DISim$_b$ case, following the same steps as for the
Maxwell-Sim case, we get
\be
L= -m  (-\eta_{\mu\nu} \dot x^\mu \dot x^\nu)^{(1-b)/2}\,
     (-n_\rho \dot x^\rho)^b\,+
   \frac12f_{\mu\nu}\left(\dot\theta^{\mu\nu}-\frac12(x^\mu\dot
x^\nu- x^\nu\dot x^\mu)\right)\,.\label{Lag5b}
\ee

\section{Hamiltonian Viewpoint}\label{hamilsec}

\subsection{Kaluza-Klein interlude}

   Before dealing with the Maxwell algebra approach,
it may be helpful to contrast the six ``angles''
$\theta_{\mu \nu}$ with the single angle $\theta$ introduced
in Kaluza-Klein approaches to motion in a homogeneous
electromagnetic field, considered as geodesic  motion in the 5-dimensional
Heisenberg group.  The Maurer-Cartan forms are
\ben
P^\mu = d x^\mu \,,\qquad
Z=d \theta - \half F_{\mu \nu} x^\mu dx ^\nu \,,
\een
with the non-trivial algebra
\ben
d Z= -\half F_{\mu \nu} P^\mu \wedge P^\nu\,,
\een
and metric
\ben
ds _5^2 = \eta _{\mu \nu} dx^\mu dx ^\nu +
 \bigl(d \theta - \half F_{\mu \nu} x^\mu dx ^\nu  \bigr)^2\,.
\label{K^2}
\een
The metric (\ref{K^2}) is invariant under the left action of the
Heisenberg group,  and an additional outer  action of the abelian
subgroup  $ G_2 \subset SO(3,1)$ generated by $\bG$ and $\bstarG$.
We may identify ${\bf Z} $ with ${\bf Z_{elec}}$ introduced earlier.
If we consider the coset (EBCR)/$(G_2,Z_{mag})$, then  the
quadratic combination $P^2+Z^2$  is invariant under the stability
group. The corresponding metric is (\ref{K^2}) and therefore it is
invariant under the whole EBCR group.

   A convenient matrix representation of the Heisenberg group is given by
\be
\begin{pmatrix} {x'}^\mu\\ \theta'\\ 1 \end{pmatrix} =
\begin{pmatrix} \delta^\mu_\nu &&&0 &&& a^\mu \\
         -\ft12 F_{\nu\lambda}\, a^\lambda &&&1&&&\alpha\\
     0&&&0&&&1 \end{pmatrix}\,
  \begin{pmatrix} x^\nu\\ \theta\\ 1\end{pmatrix}\,.
\ee

The phase or cotangent space $T^\star(G) \equiv G \times \frak{g}$ of the
Heisenberg algebra has  coordinates $(x^\mu
,\theta,p_\mu,p_\theta)$. $( \bar P_\mu, \bar M_{\mu \nu}, \bar Z) $
are the corresponding moment maps generating right actions, and
are given by
\ben
\bar P_\mu=p_\mu- \half\, p_\theta \,F_{\mu\nu} x^\nu \,,\qquad \bar
Z= p_\theta\,.
\een

 The non-vanishing  Poisson brackets of the
generators of the right actions of the Heisenberg group are
\ben
\bigl \{ \bar P_\mu ,\bar P_\nu \bigr \} = -F_{\mu \nu} \, \bar Z\,.
\een
The geodesic Hamiltonian associated to the metric (\ref{K^2}) is
\ben
H= {1 \over 2m} \bar P_\mu \bar P^\mu + {1 \over 2m}  \bar Z ^2 \,,
\label{K2ham}
\een
so that
\bea
\dot {\bar Z}  &=& 0\,,\nn\\
\dot {\bar P} _\mu &=& -{1 \over m} \bar Z   F_{\mu \nu} \bar P^\nu \,.
\eea
(Note that although the $\bar Z^2$ term in (\ref{K2ham}) is needed for
the correspondence with the metric (\ref{K^2}), {\bf however}
it plays no r\^ole in the
dynamics.)  The $x^\mu$ equation of motion is
\ben
\dot x^\mu = {1 \over m}
\bar P^\mu \,.
\een
The equation for $\theta$,  conjugate to $\bar
Z_{\mu \nu}$ is
\ben
\dot \theta  = \frac 1m \bar Z\,.
\label{Heisenberg}
\een

   The moment maps that generate {\it left} translations are given by
\be
\bar P_\mu + \bar Z\, F_{\mu\nu}\, x^\nu\,,\qquad \bar Z\,,
\ee
These are constant for a Hamiltonian such as (\ref{K2ham}), which
depends only on the
moment maps that generate {\it right} translations.

   The mechanical momentum $p_i$ and Noether momentum $P_i$ in our
discussion in the Introduction correspond to momentum maps generating
right translations and left translations respectively.

We have obtained the standard Lorentz
force equation, and the electric charge corresponds to the conserved
momentum $-\bar Z$ in the extra dimension. The externally-given
Maxwell field $F_{\mu \nu}$ is constant throughout, and is
non-dynamical. Note that we could obtain the same result with a more
general Hamiltonian of the form
\ben
H= {1 \over 2m} \bar P_\mu \bar P^\mu + \beta  \bar Z  ^2 \,,
\een
where $\beta$ is an arbitrary constant.

   We may also include a magnetic charge  by adding an extra
central extension
\ben {\starZ}=d  {\startheta} - \half {\starF}_{\mu \nu}\, x^\mu dx
^\nu \,. \een
The Hamiltonian
\ben
H= {1 \over 2m} \bar P_\mu \bar P^\mu
\een
will now lead to the constancy of both ${\bar Z}$ and $\bar
{\starZ}$ and the equation of motion
\ben
\dot {\bar P} _\mu = -{1 \over m}  \Bigl ( \bar Z   F_{\mu \nu}
+ \bar {\starZ} \starF_{\mu \nu} \Bigr )
 \bar P^\nu  \,.
\een

If we identify ${\bf \starZ } $ with ${\bf Z_{mag}} $ , then
then the six-dimensional  Heisenberg  algebra with two central charges
may be identified with the coset (EBCR)/G$_2$.
Note that  the presence of magnetic and electric charges is due to the
presence of central charges in the 8-dimensional
 EBCR algebra. These central charges
are absent  in the Maxwell algebra.

\subsection{The Maxwell algebra}

  The phase space, or cotangent space, $T^\star(G) \equiv G \times \frak{g}$
of the Maxwell algebra has  coordinates
$(x^\mu,\theta^{\mu\nu},p_\mu,f_{\mu\nu})$.
The left-invariant Maurer-Cartan forms are given by (\ref{mcformsmaxwell}) and
$( \bar P_\mu, \bar M_{\mu \nu}, \bar Z_{\mu \nu}) $ are the
corresponding moment maps generating right actions. They are given by
\ben
\bar P_\mu=p_\mu-\half\,f_{\mu\nu} x^\nu \,,\qquad \bar
Z_{\mu\nu}=f_{\mu\nu} \,,
\een

    The non-vanishing Poisson brackets are
\bea
\bigl \{ \bar M_{\alpha \beta} ,\bar P_\gamma \bigr \}&=&
\eta_{\beta \gamma} \bar P_\alpha -\eta_{\alpha \gamma} \bar P _\beta\,,\\
\bigl \{ \bar M_ {\alpha \beta}, \bar M_{\gamma \delta} \bigr \} &=&
\eta_{\beta \gamma} \bar M_{\alpha \delta} - \eta _{\beta \delta}
\bar M_{\alpha \gamma} +\eta_{\alpha \delta} \bar M_{\beta \gamma  } -
\eta_{\alpha \gamma} \bar M_{\beta \delta}\,,\\
\bigl \{ \bar M_ {\alpha \beta}, \bar Z_{\gamma \delta} \bigr \} &=&
\eta_{\beta \gamma} \bar Z_{\alpha \delta} -\eta _{\beta \delta}
\bar Z_{\alpha \gamma} +\eta_{\alpha \delta} \bar Z_{\beta \gamma  }
-\eta_{\alpha \gamma} \bar Z_{\beta \delta}\,,\\
\bigl \{\bar P_\alpha, \bar P_\beta \bigr \} &=& -\bar Z_{\alpha \beta} \,.
\eea
There are two generic Casimir functions,
\ben
C_1 = \ft12 \bigl ( \bar P^\mu \bar P_\mu + \bar M_{\mu \nu}
\bar Z^{\mu \nu} \bigr ) \,,\qquad
C_2= \half \bar Z_{\mu \nu} \bar Z^{\mu \nu}\,.
\een
For the Hamiltonian, we take
\ben
H= {1 \over 2m} \eta ^{\mu \nu}   \bar P_\mu \bar P_\nu  \,.
\een
Thus the Euler equations imply
\bea
\dot {\bar Z}_{\mu \nu} &=&0\,, \qquad \Rightarrow \qquad \bar Z_{\mu\nu} =
       - e F_{\mu\nu}= \hbox{ constant}\,,\label{em}\\
\dot{ \bar P}_\mu  &=& {1 \over m} \{\bar P_\mu ,\bar P_\nu  \}\bar  P^\nu
 = -\bar Z_{\mu \nu} \bar \, P^\nu = e F_{\mu\nu}\, \bar P^\nu\,,
\eea
and the $x^\mu$ equation of motion is
\ben
\dot x^\mu = {1 \over m} \bar P^\mu \,.
\een
The equation for $\theta ^{\mu \nu}$,  conjugate to $\bar Z_{\mu \nu}$
is
\ben \dot \theta ^{\mu \nu} =\half  (x^\mu  \dot x^\nu - x^\nu \dot
x^\mu )
   \,.
\label{Kepler}
\een
Thus  we obtain  the motion of a particle
in a constant electromagnetic field, for which the momentum vector
$\bar P^\mu(\tau)$ undergoes a constant Lorentz transformation
\ben
\bar P ^\mu (\tau)= \Big[\exp\Big({e \tau } F\Big)\Big]^\mu{}_\nu  \,
\bar P^\nu (0) \,.
\een

   By contrast with the Kaluza-Klein approach, which gives the same
equations for the $x^\mu$ variables with an externally imposed constant
Maxwell field
$F_{\mu \nu}$, in the Maxwell algebra approach we find that
the Maxwell field must be constant as a consequence of the equations
of motion. The equations for the six angles  $\theta^{\mu \nu}$
are also richer. They  may be interpreted geometrically
as follows. The curve in spacetime $x^\mu=x^\mu(\tau) $ has a  projection
onto each  $\mu$-$\nu$ 2-plane. The curve sweeps out area
at a rate
\ben
\fft{dA^{\mu \nu}}{d\tau} = \half (x^\mu \dot x^\nu - x^\nu \dot x^\mu )\,.
\een
Thus (\ref{Kepler}) may be re-written as
\ben
{d \theta ^{\mu \nu}
\over d \tau} ={d A ^{\mu \nu}
\over d \tau} \,.
\een
In other words $\theta^{\mu \nu} (\tau)$
is the total area $A^{\mu \nu} (\tau)$ swept out during the motion.

The canonical Lagrangian that reproduces the previous equation of motion
is
\crampest
\be L= \bar P_\mu\, \dot x^\mu  + \ft12 f_{\mu\nu}\,
 [\dot\theta^{\mu\nu} -\ft12 (x^\mu\, \dot x^\nu -x^\nu\, \dot x^\mu)]
-\ft12 e\, \bar P^2\,,\label{Lag2new}
\ee
\uncramp
which, apart from a constant piece, is obtained from  the
diffeomorphism-invariant
Lagrangian (\ref{Lag2}) by choosing  the proper-time gauge $e=m$.

  One may choose different Hamiltonians. For example,
\ben
H=\frac{\bar P_\mu \bar P^\mu}{2m} + \half \alpha \bar Z_{\mu
\nu} \bar Z^{\mu \nu}\,.
\een
The equations of motion are the same as before except for those of the
variables $\theta^{\mu\nu}$, which now satisfy
\ben
\dot \theta ^{\mu \nu} =\half  (x^\mu  \dot x^\nu - x^\nu \dot
x^\mu )+\alpha\,f_{\mu \nu}
   \,.\label{Keplernew}
\een

   The canonical Lagrangian
\crampest
\be
L= \bar P_\mu\, \dot x^\mu  + \ft12 f_{\mu\nu}\,
 [\dot\theta^{\mu\nu} -\ft12 (x^\mu\, \dot x^\nu -x^\nu\, \dot x^\mu)]
-\ft12 e\, \bar P^2-\frac{\alpha}{2}
f_{\mu\nu}f^{\mu\nu}\label{Lag2new1}
\ee
\uncramp
gives, after eliminating the non-dynamical field $F_{\mu\nu}$, the
Lagrangian (\ref{Lag0}).

\subsection{Other Hamiltonians}

   Those which admit a constant $\bar Z_{\mu \nu}=F_{\mu \nu}$
and are Lorentz-invariant are of the form
\ben
2 m H=\bar P_\mu \bar P^\mu + \half \alpha \bar Z_{\mu \nu} \bar Z^{\mu \nu}
- \beta \bar Z_{\mu \nu} \bar M^{\mu \nu}\,.
\een
The second term does not contribute, since it commutes with everything,
and so we drop it. Hamilton's equations then give
\ben
\dot {\bar P} _\mu  = {1 \over m}(1-\beta) F_{\mu \nu} \bar P^\nu\,.
\een
Note that in the special case $\beta=1$, we
find $\dot {\bar P} _\mu=0$.   This is not surprising,
because in that case the Hamiltonian is bi-invariant, i.e. it is a Casimir,
and hence generates no motion at all.

\subsection{Maxwell-Sim and Maxwell-DISim$_b$}

    The Maurer-Cartan forms of the coset (Maxwell-Sim)/(Sim) are the
same as in the Maxwell case (\ref{mcformsmaxwell}), and the moment maps
are also given by
\ben
\bar P_\mu=p_\mu-\half\,f_{\mu\nu} x^\nu \,,\qquad \bar
Z_{\mu\nu}=f_{\mu\nu} \,.
\een
The geodesic Hamiltonian is given by
\ben
H=\frac{\bar P_\mu \bar P^\mu}{2m} + \half \alpha \bar Z_{\mu
\nu} \bar Z^{\mu \nu}\,,
\een
and therefore reproduces the same dynamics as in the Maxwell case.

For the case of the coset (Maxwell-DISim$_b$)/(Sim), the
Maurer-Cartan forms and the momenta are the same as for the Maxwell
case.
%
%

\subsection{Hamiltonian Treatment of the Bogoslovsky-Maxwell algebra}

  In previous work \cite{gibgompop} we obtained a  Finslerian Lagrangian
invariant under DISim(2)$_b$, where $b$ is the deformation parameter
constructed from the Finslerian line element
\ben
ds^2= -F(v^\mu)^2 \, d\tau^2\,,
\een
where the Finsler function $F(v^\mu)$ is
homogeneous of degree 1 in the four-velocity
$v^\mu= dx^\mu/d\tau$.  In general, if we were to use a
multiple of the Finlser function $F(v^\mu)$
as a Lagrangian $L(v^\mu$), then its Legendre transform would vanish,
since a  Lagrangian  which is homogeneous of degree $k$ in velocities
gives, on taking a Legendre transform, a Hamiltonian
\bea
H(p_\mu) &=&  v^\mu p_\mu  - L (v^\mu) \,,\\
& =& v^\mu \frac{\p L }{\p v^\mu} -L \\
         &=& (k-1) L \,,
\eea
which is homogeneous of degree $ \frac{k}{k-1}$ in  momenta $p_\mu$ .
If $k=2$  we have
\ben
H(p)=L(v) \,,
\een
and both are of degree two. Therefore it is customary in Finsler
geometry to set
\ben
L(v^\mu) = -\half m F^2(v^\mu) \,.
\een
For the case of Bogoslovsky's Finslerian geometry we would then  have
\ben
L= -\half m\, (-n_\rho v^\rho) ^{2b} (-\eta _{\mu \nu} v^\mu v^\nu) ^{1-b} \,,
\label{Laga}
\een
where $ n^\mu= \eta ^{\mu \nu} n_\nu $ is a constant
future-directed null vector.
The minus signs appear in (\ref{Laga}) because $v^\mu$ is assumed
to be future-directed and timelike.  With our signature convention,
the inner product $n\cdot v=n_\mu v^\mu$ is then negative.
We find that
\ben
p_\mu= b m n_\mu (-n\cdot v) ^{2b-1} (- v^2   )^{1-b}
+ (1-b) m v_\mu (-n\cdot v) ^{2b} (-v^2 )^ {-b}
\,,
\een
and
\ben
H= -\fft1{2m}\, \Bigl( - \frac{p^2 }{1-b^2}  \Bigr)
^{1+b}  \Bigl(  -\frac{n\cdot p }{ 1-b} \Bigr)^{- 2b }
\,.\label{Ham}
\een
Imposing the mass-shell condition $H=-\ft12 m$, i.e. $F(v)^2=1$,
leads to equation (18) of
\cite{gibgompop}.  In this case, the parameter $\tau$ coincides with the
Finslerian measure of proper time along the world-line of the
particle.\footnote{We could instead impose the gauge condition $v^2=-1$,
but this is less natural in the Finslerian framework.}
Equation (\ref{Ham}) is also equivalent to the expression
(\ref{mass2}) (with $\alpha=-m(1-b)$ and $\beta=-\ft12 m(1+b)$).

   We may also give the expression for $v^\mu=\del H/\del p_\mu$, finding
\be\label{velocity}
 v^\mu = e\Big[ p^\mu - b\, m\, \Big(-\fft{n^\nu
p_\nu}{m(1-b)}
        \Big)^{\fft{b-1}{b+1}}\, n^\mu \Big]\,,
\ee
where
\be
e= -\fft{(1+b)m}{p^2}\,.
\ee
Again, this is in agreement with the corresponding expression (\ref{vexp})
obtained in the Lagrangian treatment.

   The Lorentz force equation follows from (\ref{velocity}), and
$\dot{\bar P}^\mu=
\{\bar P^\mu, H\}$, which implies
\be
\dot{\bar P}^\mu = -\bar Z^{\mu\nu}\, v_\nu\,.
\ee

\section{Conclusions}

   We have constructed the non-central extensions and deformations of the
ISim algebra. The Maxwell-Sim algebra is obtained from the
translation generators $\bP_\mu$  and  the non-central extension $
\bZ_{\mu\nu}=  [\bP_\mu,\bP_\nu]$, together with the Sim$(n)$
generators $(\bM_{+i}, \bM_{+-},\bM_{ij})$.

  In general dimensions, the deformations of Maxwell-Sim algebra are
characterised by two parameters $b$ and $c$. The deformation parameterised by
$c$ is the analogue of the $k$ deformation of the Maxwell algebra found in
\cite{gomkamluk}, which gave $SO(3,2)\times SO(3,1)$
or $SO(4,1)\times SO(3,1)$ depending on the sign of $k$.
The $b$-deformation of
Maxwell-Sim produces the Maxwell extension of the DISim$_b$ algebra, which
is related to Finslerian geometry.

 We have also studied the motion of a massive particle
interacting with a constant electromagnetic field with these
symmetries. In the case of Maxwell-DISim$_b$, the motion is given by
a Finslerian Lorentz force, whilst by contrast for the undeformed
Maxwell-Sim algebra we obtain the ordinary Lorentz force.

\section{Acknowledgements}

    We acknowledge discussions with Sotirios Bonanos, Roberto
Casalbuoni, Kiyoshi Kamimura, David Kubiznak and Mikhail Vasiliev.
This work was initiated at the Galileo
Galilei Institute in Florence, continued at the Benasque School, at
CTC in DAMTP  and concluded in the Department of Physics in
Barcelona. The authors would like to thank those institutions for
their hospitality and support. We also acknowledge financial support
from projects FPA2007-66665-C02-01, 2009SGR502 and Consolider CPAN
CSD2007-00042.  The work of C.N.P. is supported in part by
DOE grant DE-FG03-95ER40917.

\appendix

\section{Conventions}

In this appendix we record some of our
conventions and notation  when working with Lie groups,
Lie algebras and Poisson algebras.

  Given a Lie group $G$, with coordinates $x^\mu$, i.e. group elements
$G \ni g=g(x^\mu)$, and  left and right invariant Cartan-Maurer forms
\ben
 g^{-1} dg= \lambda ^a {\bf e}_a  \,,\qquad  dg g^{-1} =
\rho ^a {\bf e}_a\,,\label{lamrhodef}
\een
with ${\bf e}_a$ a basis for the Lie algebra $\frak{g}$
such that
\ben
[{\bf e}_a ,{\bf e}_b ] =C_a\,^c\,_b  \,{\bf e}_c \,,
\een
the Maurer-Cartan equations are
\ben
d \lambda ^c = -\half C_a\,^c\,_b \,\lambda ^a \wedge \lambda ^b
\,,\qquad d \rho ^c = \half C_a\,^c\,_b \,  \rho ^a \wedge \rho ^b \,.
\een

   The left and right  invariant
vector fields $L^\mu_a$ and $R^\mu _a $ dual to
$\lambda^a _\mu$ and $\rho ^a _\mu $  respectively,
\ben
\lambda^a _\mu L _b^\mu =
 \delta ^a_b \,,\qquad
\rho^a _\mu R _b^\mu = \delta ^a_b \,,
\een
satisfy
\ben
[L_a, L_b ] = C_a\,^c\,_b \, L_c
\qquad [R_a, L_b ]=0\,, \qquad [R_a, R_b ] = -C_a\,^c\,_b \,  R_c \,,
\een
and respectively generate  right  and left translations on $G$.

   Quantum mechanically, one often inserts $i$'s so that
if $\hat R_a ={1 \over i} R_a$, $\hat L_a ={1 \over i} L_a   $  then
\ben
[ \hat R_a, \hat R_b ] = i C_a\,^c\,_b \, \hat R_c\,,
\een
\ben
 [ {\hat L} _a, {\hat L}_b  ]   = -i C_a\,^c\,_b \,{\hat L} _c \,.
\een
The $\hat R_a$ and $\hat L_a$ vector fields are then operators acting
on complex-valued  functions of the group coordinates $x^\mu$.

   Thinking of $G$ as a configuration space,
we can pass to the phase space or cotangent space $T^\starG \equiv G \times
\frak{g}$, with
coordinates $(x^\mu ,p_\nu)$.
The actions of $G$ on $G$ then lift to  $T^\starG$ as
canonical transformations,  leaving the natural symplectic form
$d p_\mu \wedge dx ^\mu$ invariant. Given the symplectic form,
we can introduce the Poisson bracket as usual. In local Darboux
coordinates $( x^\mu , p_\nu )$, it is given by
\ben
\{ f,g\} = {\p f \over \p x^\mu} {\p g \over \p p_\mu } -
 {\p g \over \p x^\mu}  {\p f \over \p p_\mu}\,,
\een
so that
\ben
\{ x^\mu, p_ \nu \} = \delta ^\mu _\nu \,.
\een
Infinitesimally, the lifts of left and right actions are canonical
transformations
generated by ``generating functions''  or ``moment maps.''
Because, in general,  we have both left and right actions to take into
account,  we define two sets of moment maps into $\frak{g}^\star $, the dual
of the Lie algebra,
\ben
M_a= p_\mu L^\mu _a\,,\qquad N_a= p_\mu R^\mu _a\,,
\een
with Poisson brackets which are readily seen to be
\ben
\{M_a, M_b \} = -C_a\,^b\,_c M_b
\qquad \{M_a, N_b\}=0\,, \qquad \{N_a, N_b \} = C_a\,^b\,_c N_b \,.
\een
The moment maps $M_a$  generate the lifts of right  translations
and the moment maps $N_a$ generate the lifts of left translations.

   A Hamiltonian $H=H(x^\mu, p_\mu)$,  which is left-invariant,
satisfies
\ben
\dot N_a = \{N_a , H \}=0\, ,
\een
and so the moment maps $N_a$ are constants of the motion.
By contrast, the moment maps $M_a$ generating right actions
are time-dependent,
\ben
\dot M_a = \{M_a , H \} \ne 0\,.
\label{Euler}
\een

A left-invariant Lagrangian may be constructed from
combinations of left-invariant velocities or angular velocities
\ben
\omega^a =\lambda ^a _\mu  \dot  x ^\mu\,.
\een
Thus the Hamiltonian is a combination of
the momenta maps $M_a$,
\ben
H= H(M_a) \,.
\een
Thus (\ref{Euler}) provide an autonomous 1'st-order  system
of ODEs on $\frak{g}^\star$ for the moment maps
$M_a$, called the {\it Euler equations}.
To obtain the motion on the group, one uses the
equation
\ben
\dot x ^\mu = {\p H \over \p p_\mu}\,.
\een
Now
\ben
p_\mu = M_a \lambda ^a _\mu \,,
\een
and so
\ben
\dot x^\mu = L_a ^\mu {\p H \over \p M_a} \,.
\een
\section{Lifshitz and Schr\"odinger algebras}

In this appendix we shall describe the connection between
the deformed inhomogeneous Sim algebra $\frak{disim}_b(k)$
and the Lifshitz,  Schr\"odinger
and extended Schr\"odinger algebras,
 $\frak{lif}_z$, $\frak{sch}_z(k)$ and $\widetilde{\frak{sch}}(k)$ respectively.
We start with

\subsection{Lifshitz scaling}

In non-relativistic theories with $k$ spatial dimensions,
one is interested in the behaviour of physical quantities
under  what has come to be called {\it Lifshitz scaling}, i.e.
under
\ben
 t \rightarrow \lambda ^z t \,,\qquad {\bf x} \rightarrow \lambda {\bf x}
\een
where $t$ is the time variable and ${\bf x}= (x_1,x_2,\dots,x_k )$ is
the spatial position vector.

 If $D$ generates scalings or  dilatations we may combine this
with space translations
$P_i$, spatial rotations, $M_{ij}$ and
time translations $H$,  to obtain the  {\it Lifshitz  Algebra},
$\frak{lif}_z(k)$
in $k$ spatial dimensions,
\ben
\bigl [D,M_{ij} \bigr ]= 0\,,\qquad \bigl [D,P_i\bigr ]=P_i
\,,\qquad  \bigl[ D,H \bigr ]
=z H \,,
\een
 where the obvious  brackets for $M_{ij}$ have been omitted.
The Lie  algebra
spanned by $D$,
 $P_i$, and $H$  is therefore invariant under the adjoint  action of
the rotation subalgebra $\frak{so}(k)$  generated by $M_{ij}$.
If $i=1,2,\dots ,k$, then  $\frak{lif}_z(k)$  has dimension $\half k(k+1) +2$
and the quotient $\frak{lif}_z (k)/ \frak{so}(k)$  has dimension $ k+2$.

\subsection{Lifshitz spacetime}

This is a $k+2$ dimensional spacetime  equipped  with
a metric invariant under the left action of
the $(k+2)$-dimensional group generated by
$P_i$, $H$ and $D$.  A  Maurer-Cartan basis for this solvable group is
\ben
e^r ={dr \over r}\,,\quad  e^i ={d x^i \over r} \,, \quad
e^0= {dt  \over r^z} \,.
\een
The Lifshitz metric is then
\ben
ds_{k+2}^2 = L^2 \Bigl \{ -{dt ^2 \over r^{2z }} + {d x_i dx_i \over r^2 }
+ {d r ^2 \over r^2 }   \Bigr \}\,,
\een
with Killing vector fields corresponding to
\ben
M_{ij}=-(x_i \p_j-x_j \p_i)\,,\quad  P_i= -\p_i\,,\quad H= -\p_t\,,\quad
D= -( z t \p_t + x_i \p_i + r \p_r) \,.
\een

\begin{itemize}
\item As $r \rightarrow \infty$ we approach a singular horizon (IR limit) .
\item As $r \rightarrow 0$ we approach infinity (UV limit)
\end{itemize}

The boundary metric at infinity is obtained by taking out
a factor of $r^2$ and letting $r \rightarrow 0$:
\ben
ds_{k+2}^2 = {L^2 \over r^2 }
 \Bigl \{ -{dt ^2 \over r^{2(z-1)  }} + {d x_i dx_i  }
+ {d r ^2  }   \Bigr \}
\een

Thus
\ben
 ds^2_{\rm boundary}\,=d x_i dx_i - r^{2 (1-z)} dt ^2 \,,
\een
the speed is $c(r)= r^{(1-z)} $, and
\begin{itemize}
\item If $z>1$, we obtain infinite speed
(the boundary lightcone opens out to a plane)
\item If $z=1$, we obtain finite speed (the boundary lightcone
remains a cone)
\item If $z<1$, we obtain zero  speed (the boundary lightcone closes up
to a half line )
\end{itemize}
Strictly speaking, in the $z>1$ case,
  we need to consider the inverse metric when taking
the limit $r \rightarrow 0$.

\subsection{The boost-extended Lifshitz algebra}

One  may extend the Lifshitz algebra to include boosts
$K_i$. The scaling dependence of $K_i$ is then determined  by
its commutation relations.  Since $K_i$ is a vector we have
\ben
\bigl [ K_k, M_{ij} \bigr ] =
 - \bigl ( \delta_{ki}K_j-\delta_ {kj} K_i \bigr)
\een
For the {\bf  {\it Galilei}}   group,
\bea
\bigl [ K_i, P_j \bigr ]&=& 0\,,\\
\big [ K_i , H  \bigr ] &=&  P_i \,,
\eea
which implies that we must take
\ben
\bigl[ D, K_i \bigr ]=  ( 1-z) K_i \,.
\een
For the {\it  Carroll}  group
\bea
\bigl [ K_i, P_j \bigr ]&=& \delta_{ij} H\,,  \\
\big [ K_i H  \bigr ] &=& 0 \,,
\eea
which implies that we must take
\ben\bigl[ D, K_i \bigr ]=  ( z-1) K_i \,.\een

In the case of the Poincar\'e group
there is no choice, and one must
take  $z=1$.

\subsection{${\rm DISim} _b (k)$ }

Recall that $ {\rm DISim} _b(k) $ is a  deformation of the ${\rm ISim}(k)$
subgroup of the Poincar\'e group in $(k+2)$  spacetime
dimensions,  depending on a parameter $b$,
which may be regarded as a subgroup of the inhomogeneous  {\it Weyl group} or
{\it Causal group}
(i.e the semi-direct product of   Poincar\'e with dilatations),
in which the actions of a boost and dilations are identified
up to a factor
\cite{gibgompop}.
It is thus of dimension $\half k(k+1) + k + 3$.

If the translations are $P_+,P_-, P_i$, and the boosts are
$M_{+i},\, M_{+-}$, 
then   the non-trivial Lie brackets  are given
by
\bea
&&[M_{+-},P_\pm]= -(b\pm 1) P_\pm\,,\quad [M_{+-} ,P_i]=-b P_i\,,\nn\\
&&[M_{+-},M_{+ i}] = -M_{+ i}\,, \quad [M_{ + i}, P_-]=P_i\,,\nn\\
&&[M_{ + i},P_j] = -\delta_{ij} P_ + ,
\eea
The $\frak{so}(k)$  rotations  have the standard brackets
and act on $P_i$ and $M_{+i}$ as vectors.
The boost generator $M_{+-}$  acts on $(k+2)$-dimensional Minkowski spacetime
as
\ben
x^i\rightarrow \lambda^{-b}\, x^i\,,\quad
x^-\rightarrow \lambda^{1-b}\, x^-\,,\quad
x^+\rightarrow \lambda^{-1-b}\, x^+\,. \nn
\een
If $b=0$, then $M_{+-} $ acts as an ordinary boost..

\subsection{The Schr\"odinger and Extended Schr\"odinger algebras}

In $k$ spatial dimensions, the centrally extended  $(\half k(k+1) + k +3)$
dimensional
{\it Schr\"odinger algebra}
(in current terminology \cite{balasubra}),
which we denote $\widetilde {\frak{sch}} _z(k)$,
is obtained by adjoining Galilean boosts $K_i$, and a central term $N$
to the {\it Aristotelian algebra}
of  translations, rotations and time translations, such that
\bea
\Bigl[ M_{ij}, K_k  \Bigr ]&=&
  \bigl (\delta_{ik} K_j - \delta _{jk} K_i \bigr )\,,\\
\Bigl[P_i, K_j  \Bigr ]&=&- \delta_{ij}N\,, \\
\Bigl[H, K_i  \Bigr ]&=&- P_i \,.
\eea
The  result is the $(\half k(k+1) + k +2)$ dimensional
{\it Bargmann algebra}, a central
extension of the $(\half k(k+1) + k +1)$ dimensional
{\it Galilei} algebra. One then adjoins
a dilatation $D$,
\ben
\Bigl[D, K_i  \Bigr ] = (1-z) K_i \,,\qquad
\Bigl[D, N \Bigr ] = (2-z) N \,.
\een

If $k=3$ this is 12-dimensional,
whereas what has been called the Schr\"odinger group, i.e.
the conformal symmetry group of the  free Schr\"odinger equation
(corresponding
to $z=2$)   is 13-dimensional\footnote{The reader should note the difference with the Galiean Conformal algebra obtained by contraction from the relativistic conformal algebra, which has 15 generators (see, for example, 
\cite{negroolmo} \cite{Lukierski:2005xy}).}.  This is because the special
conformal or temporal inversion operator has been left out.

One may consistently drop the central extension
 $N$  from the Bargmann algebra to get the Galilei algebra,
and then the extended Schr\"odinger algebra $\widetilde{ \frak{sch}}(k)$
reduces to the $(\half k(k+1) + k+2)$ dimensional
unextended Schr\"odinger algebra $\frak{sch}(k)$.
If one then drops the boost generator $K_i$
one gets the Lifshitz algebra  $\frak{lif}_z(k)$.

It is  well known that  non-relativistic symmetries
and non-relativistic
conformal symmetries (Schr\"odinger algebras) in $k$ spatial
dimensions may be thought of
as subgroups
of relativistic or conformal symmetries in $k+2$
dimensional Minkowski  spacetime which commute with light-like translations.
Thus it is no surprise that
\ben
\widetilde { \frak{sch}} _z(k) \equiv \frak{disim}_b(k) \,,
\qquad b=\frac{1}{1-z} \,. \label{correspondence}
\een
To see this, one must identify the generators as follows;
\ben
H\leftrightarrow P_-\,,\qquad N\leftrightarrow-P_+ \qquad P_i
\leftrightarrow P_i\,,\qquad K_i\leftrightarrow  M_{+i}  \,.
\een
and
\ben
D\leftrightarrow (z-1) M_{+-} \,.
\een

Note that it is also possible to obtain the Lifshitz algebra $\frak{lif}(k)$
as a truncation of the $\frak{disim}_b(k)$  algebra by
discarding the $P_i$ generators and
making the identifications
\ben
H\leftrightarrow P_-\,,\qquad  \qquad P_i
\leftrightarrow M_{+i} \,,\qquad D \leftrightarrow M_{+-}\,,
\een
and
\ben
z=(b-1) \,.
\een
However this is perhaps less useful than the identification
(\ref{correspondence}).

\subsection{ Schr\"odinger spacetime}
This is $(k+3)$-dimensional, and has metric
\ben
ds ^2_{k+3} = L^2 \Bigl \{
 - {dt ^2 \over r ^{2 z} } - { 2dt dv\over r^2}
+ {d x_i dx_i \over r^2 } +{dr ^2 \over r ^2   }
\Bigr     \}
\een
with Killing vectors
\ben
K_i =-(t\p_t + x^i \p_v) \,,\qquad N= - \p_v\,,
\een
\ben
P_i =-\p_i\,,\qquad M_{ij} = - ( x_i \p_j-x_j \p_i) \qquad
D= - (z t \p_t + x_i \p_i +(2-z) v \p_v + r \p_r )\,.
\een

A Cartan-Maurer basis for this solvable group manifold is given by
\ben  e^r =dr/r\,, \quad  e^i =  {dx^i \over r} \,,
\quad e^v= {dv \over    r^{2-z}}\,, \quad
e^ t = {dt \over r^{z} } \,.
\een

\subsection { Lifshitz spacetime  as a null reduction of Schr\"odinger
spacetime}

If we identify points in the Schr\"odinger
spacetime under the ${\Bbb R}$  action generated  by the
the null Killing field $\p_v$ , i.e., under the action
of the ``central''  element $N$, we obtain
the Lifshitz spacetime. On the boundary we have the metric
\ben
ds^2_{\rm boundary} = dx _i d x_i  -2 dt dv - r^{2 (1-z) } dt ^2\,.
\een

In the cases $z >1$, we may regard the boundary as  the $(k +2)$-dimensional
Duval-Kunzle spacetime whose null reduction produces
the $(k+1)$-dimensional  Newton-Cartan spacetime.
Strictly speaking we need to consider the inverse metric when taking
the limit.

\end{document}